 \documentclass{camera}
\usepackage{psfig}
\usepackage{epsfig}
\usepackage{graphicx}

\def \deg {$^\circ$}

  \renewenvironment{thebibliography}[1]{%
    \begin{oldthebibliography}{#1}%
      \setlength{\parskip}{2mm}%
      \setlength{\itemsep}{0ex}%
  }%
  {%
    \end{oldthebibliography}%
  }

\begin{document}

\title{POSITRONS IN THE GALAXY: THEIR BIRTHS, MARRIAGES AND DEATHS.}


\author{GERALD K. SKINNER}

%
\organization{NASA-GSFC / CRESST / University of Maryland,\\
Code 661 NASA-GSFC, Greenbelt, MD 20771, USA}

\maketitle

\begin{abstract}
High energy ($\sim$GeV) positrons are seen within cosmic rays and observation of a narrow line at 511 keV shows that positrons are annihilating in the galaxy after slowing down to $\sim$keV energies or less.
Our state of knowledge of the origin of these positrons, of the formation of positronium `atoms' , and of the circumstances of their annihilation or escape from the galaxy are reviewed and the question of whether the two phenomena are linked is discussed.  

\end{abstract}
\vspace{1.0cm}

\section{Introduction}

It is convenient to divide positrons in the galaxy into two categories. High energy ($\sim$GeV) positrons are observed directly in cosmic rays.   They represent roughly 10\% of the Leptonic content of primary cosmic rays, though this fraction is somewhat energy dependent.
Lower energy positrons are seen indirectly -- the observation of a gamma-ray line at 511 keV shows  that positrons are  annihilating with electrons and the narrowness of the line is an indication that annihilation takes place when the particles have relatively low energy. The positrons responsible for the line are thought to have slowed down to energies $\sim$keV from initial energies  that may have been in the MeV region, or perhaps higher. 

Each of these two groups, and whether or not they are related,  will be discussed here, but as the former are considered  in other papers in this volume (Bossi, 2010; Bruno, 2010; Regis, 2010) emphasis is on what can be deduced  of  low energy positrons  from observations of 511 keV gamma-ray and of the associated positronium continuum radiation. 

\section{High Energy Positrons}

Our knowledge of high energy positrons in the galaxy is here summarized briefly. For more information see the papers cited above and references therein.

Anderson (1932) first discovered  positrons  among the {\it secondaries} of atmospheric cosmic ray showers. Much later,  positrons were observed using balloon-borne instruments in the cosmic rays incident on the top of the earth's atmosphere (Fanselow, 1969, Buffington, 1974;  Daugherty, 1975). The positron fraction  was about 0.1, consistent with expectations assuming they were  produced in interstellar space by cosmic rays encountering $4~g~cm^{-2}$ of material.
  
Even in these early measurements there were hints of a deviation above a few GeV from the gradual decline of positron fraction with increasing energy expected if the positrons originated in collisions of cosmic rays with  interstellar matter.  Such a deviation has now been  seen with high significance in the results obtained with the PAMELA instrument   (Adriani et al., 2009). It has been tentatively associated in the literature with  excesses, compared with predictions, in the {\it total} $(e^++e^-)$ flux at 100--1000 GeV seen  with the Fermi LAT (Abdo,et al., 2009) and with HESS (Aharonian et al., 2009). Links have been proposed between both excesses  and  the microwave `haze' seen around the galactic center with WMAP (Finkbeiner, 2004, and references therein) and a similar one seen in high energy gamma-rays with Fermi/LAT (Dobler et al, 2009). 

This combination of unexpected phenomena, quite possibly related to each other, has led to widespread suggestions  that they are associated with the annihilation, or perhaps decay,  of dark matter (e.g. Salati, 2010; Arkani-Hamed, 2009, but also several hundred other recent papers).     On the other hand it has also been argued that the particle excesses can be explained as due to nearby pulsars (Profumo, 2008, Hooper et al., 2009).  The WMAP `haze'  could then be due to spinning dust  while the Fermi one could be due to a combination of other components and systematic effects (Linden and Profumo, 2010).

It is natural to consider whether  the high energy cosmic ray positrons could the the same ones that slow down to annihilate at low energies and produce the 511 keV gamma-ray line and that are the main subject of this review. Such a connection has been suggested by (Ramaty et al., 1970),  but it will be argued in \S\ref{link} that this is unlikely. 

\section{Low Energy Positrons} 

\subsection{History}

Early observations with balloon-borne instruments showed a gamma-ray line from the general direction of the galactic center. The line at first appeared, using  scintillation detectors, to be broad and at an energy that initially seemed  to exclude it being due to  511 keV positron annihilation radiation. Johnson, Harden and Hames (1972) found  473$\pm$30 keV, while with additional data Johnson and Haymes (1973) reported 485$\pm$35 keV.  Using a Germanium detector with very much better energy resolution, Leventhal et al. (1978) finally showed that the energy of the line is indeed 511 keV. 

\begin{figure}
   \begin{center}
   \begin{tabular}{c}
 \includegraphics[trim = 0mm 15mm 0mm 33mm, clip, width=12cm]{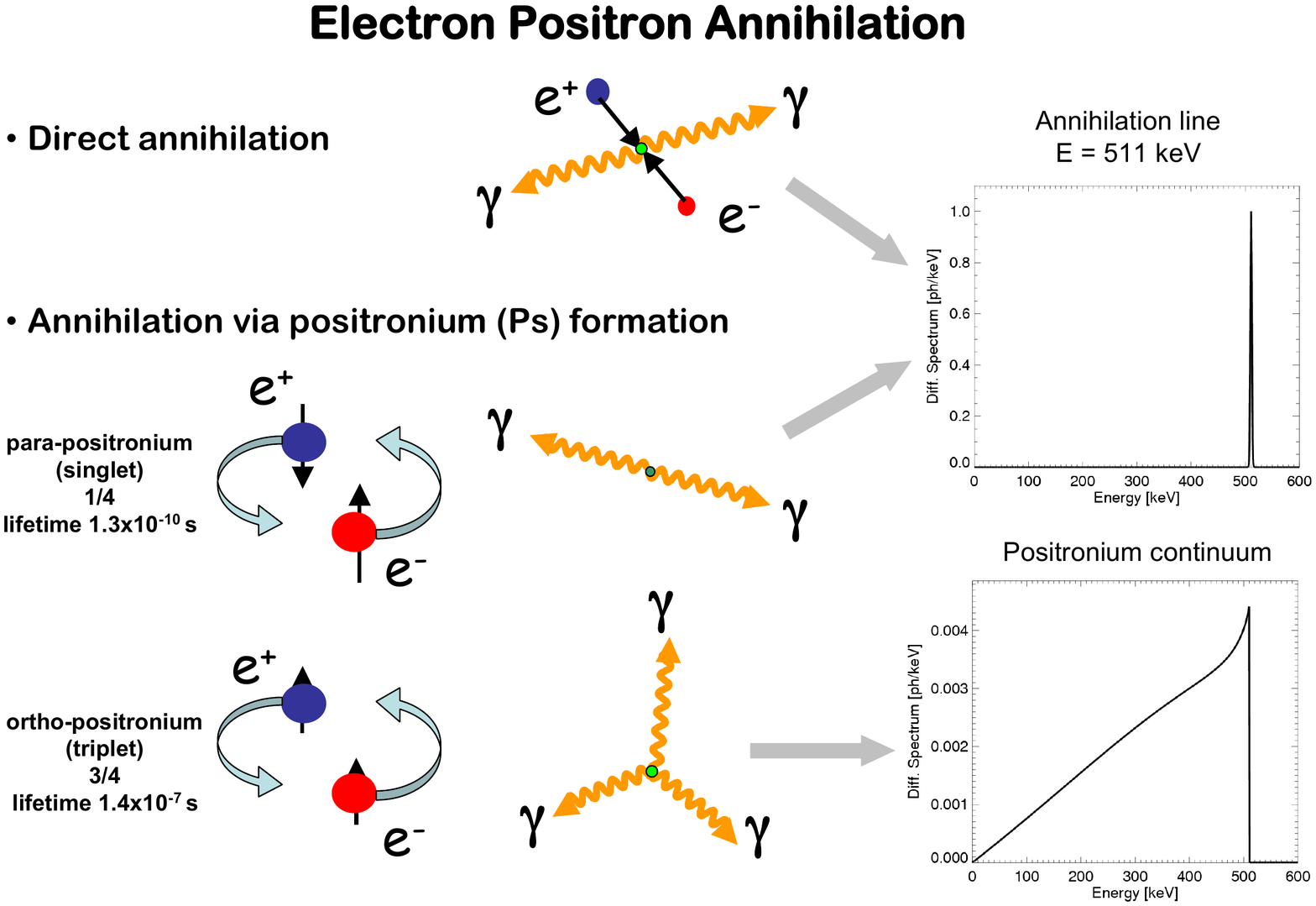}\\
   \end{tabular}
   \end{center}
   \vspace{-8mm}
   \caption[example] 
   { \label{fig:decay_schemes}  Paths for the annihilation of positrons with electrons.}
   \end{figure}

Annihilation of positrons can take place either directly, leading to the production of two opposed 511 keV photons, or via the formation of a positronium `atom'. In the latter case, if the spins of the two electrons are parallel (para-positronium) the state is short-lived and  decays, again producing two 511 keV photons.  3 times out of 4, however, the spins are anti-parallel (ortho-positronium) and 2-photon decay is forbidden, in which case three photons are produced, with a continuum of energies up to a maximum of 511 keV (Fig. \ref{fig:decay_schemes}).

Leventhal et al. showed that the 3-photon continuum was present in the galactic emission at a level that implied that 
a large fraction of annihilations take place via the formation of positronium. This continuum may explain 
the fact that estimates of the line energy from earlier, low resolution, measurements  were biased towards lower energies.
It gradually became clear that apparent variability in  early measurements of  the 511 keV line flux was due to a combination of  systematic errors and  comparing observations made with instruments having different fields of view  where the source is in practice extended (e.g. Share et al., 1990).

 The observations that first started to delineate the form of the extended emission were those made with the OSSE instrument on the Compton-GRO observatory. Although OSSE did not have the fine energy resolution of Germanium detectors, it was able to confirm that the brightest 511 keV emission is from an extended `bulge' region around the galactic center while the surface brightness of a second component that  extends along the galactic plane is much lower. Evidence for a so-called `Positive Latitude Enhancement'  (Cheng et al. 1997; Purcell et al. 1997) became less convincing when systematic errors were better understood (Milne et al, 2000).

\subsection{Recent Results}

The most detailed information on the annihilation radiation now comes from the SPI instrument on the INTEGRAL observatory which has allowed for the first time observations of the annihilation radiation with the high spectral resolution  of Germanium detectors and with simultaneous imaging. 

\begin{figure}
   \begin{center}
   \begin{tabular}{c}
 \includegraphics[trim = 0mm 0mm 0mm 0mm, clip, width=8cm]{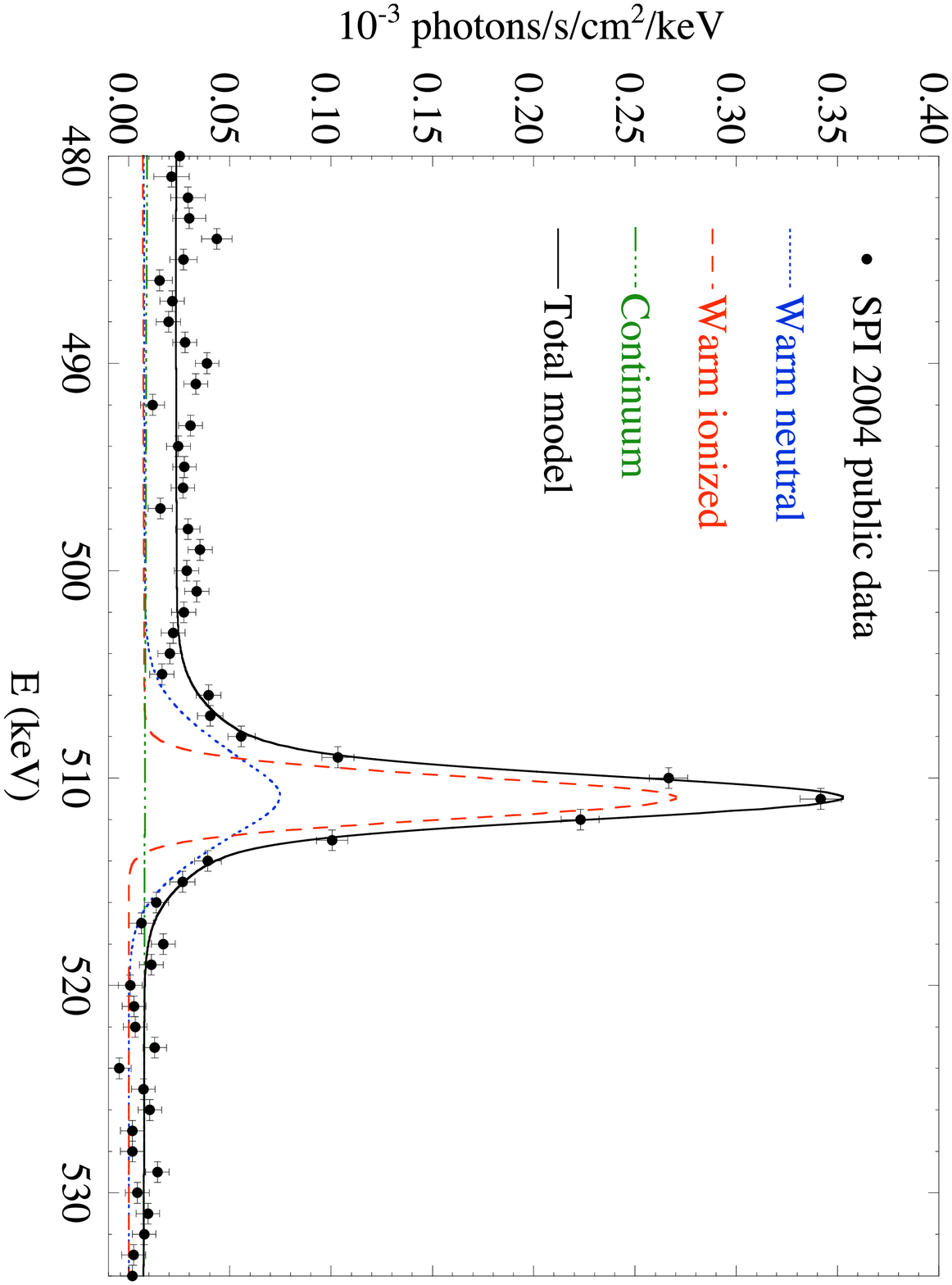}\\
   \end{tabular}
   \end{center}
   \vspace{-8mm}
   \caption[example] 
   { \label{fig:spectrum}   Interpretation of the  shape of the 511 keV line in terms of components from the warm ISM and  Galactic continuum emission (Jean et al. (2006).}
   \end{figure}

All of the possible production mechanisms lead to positrons with an initial energy of $\sim$MeV or more, so the fact that the 511 keV line is relatively narrow means that the positrons producing this line must have lost most of their energy before annihilating. It is not excluded, however, that some `inflight annihilation' takes place -- the Doppler shifted radiation would form a continuum over a wide range of energies that might not be discernible amid other such radiation.   The 511 keV line, though narrow,  is resolved and its width and shape, together with the fraction of photons in the 3-photon continuum,  allow deductions to be drawn about the circumstances and the environment in which annihilation takes.
Churazov et al. (2005) find that the annihilation is predominantly in the warm, partly ionized, ISM, while Jean et al. (2006) obtain a best fit with a mixture of warm ionized and warm neutral phases (Fig. \ref{fig:spectrum}). Significant contributions from annihilation in the hot ISM or in the cores of molecular clouds are excluded.

The SPI results on the sky distribution of the 511 keV line (see Weidenspointner et al., 2008) may be summarized as follows :

\begin{itemize}
\item {Images of the galaxy in the 511 keV line are dominated by a central bulge whose form can be characterized as the combination of two Gaussian functions with FWHM about 3.5\deg\ and  11.5\deg\ respectively.}
\item {Consistent with OSSE results, the bulge-to-disk ratio, though poorly defined because of uncertainty in the latitude extent and any possible halo component,  is surprisingly high }.
\item{} The emission from the inner galaxy appears to exhibit  an unexpected asymmetry, with the line at negative longitudes ($-$50\deg $<l<$  0\deg ) being $\sim$1.8 times stronger than that at corresponding positive longitudes.
\end{itemize}

Like the line shape, the sky distribution of the radiation tells us about where the annihilation, the {\it death }, of the positrons takes place. Depending on how far  they travel before annihilation, discussed below, it may or may not reflect where they are produced and the question of their {\it birth} remains open. The difficulty is not that there is no explanation of how they could be produced but that there are too many. Skinner et al. (2009) divided the origins that have been proposed into 18 categories, many of them having been suggested in many different variants. Certain processes must certainly contribute at some level. For example the  decay  of  $^{26}$Al atoms, that leads to the observed 1809 keV  $^{26}$Mg gamma-ray line from the  de-excitation  of the daughter  $^{26}$Mg$^*$, produces a positron. Because of their  $\sim10^6$ y  half-life  there is ample opportunity for the $^{26}$Al atoms to escape their birth site into regions of low enough density that gamma-ray  photons can reach us.  

On the (perhaps over-simplistic) assumption that the distribution of annihilation of positrons reflects the location of their production,  the high 511 keV bulge-to-disk ratio argues against the dominant source being processes, such as $^{26}$Al decay, that occur largely in the disk. Indeed it is quite difficult to account for such a central concentration, which  has led to  many of the suggested production mechanisms involving annihilations or decay of `Dark Matter'. The observed asymmetry in the 511 keV emission  has been taken as evidence against such theories.

   \begin{figure}
   \begin{center}
   \begin{tabular}{c}
 \includegraphics[trim = 0mm 0mm 0mm 0mm, clip, width=9cm]{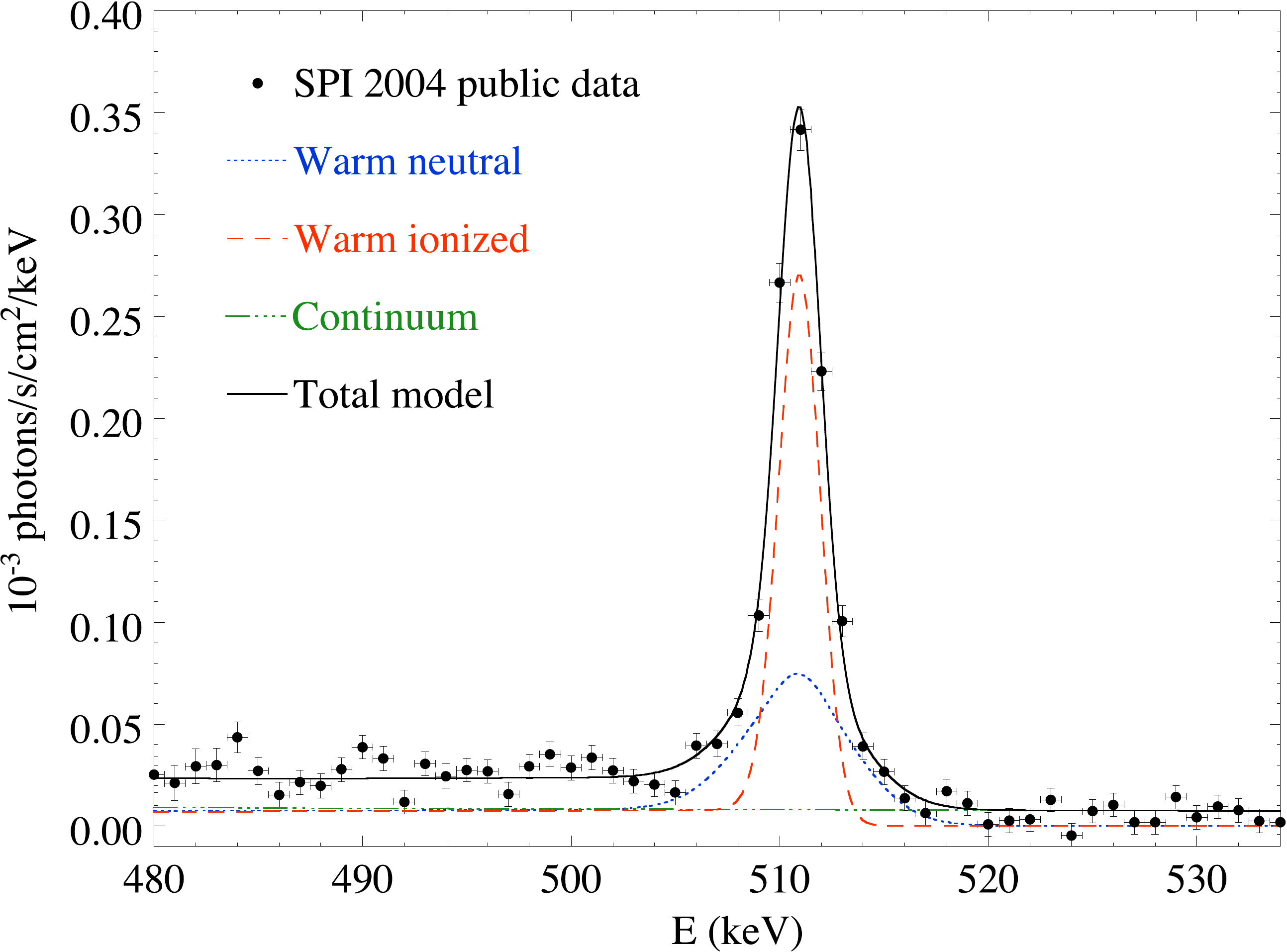}\\
   \end{tabular}
   \end{center}
   \vspace{-5mm}
   \caption[example] 
   { \label{fig:image}    The sky distribution of the 511 keV gamma-ray line radiation, observed with the INTEGRAL/SPI  instrument.}
   \end{figure} 

Weidenspointner et al. pointed out the similarity between the asymmetry and  an equally unexplained one in the distribution of Low Mass X-ray Binaries (LMXBs) observed at hard X-ray energies. It was emphasized (see Skinner et al., 2009) that the  significance of the asymmetry in  LMXBs  is much less than that of the asymmetry in the 511 keV radiation. Indeed, the similarity may be coincidental as argued by Bandyopadhyay et al. (2009), though it is interesting that more recent surveys of LMXBs  continue to show much the same effect (Table \ref{table:asym}) that was first noticed in the 3rd IBIS catalog.

\begin{table}[htdp]
\caption{The asymmetries in the number of LMXBs in different surveys compared with  that seen in the 511 keV radiation. Probabilities based on the likelihood according to the Binomial distribution of a distribution as unbalanced as that seen are expressed as an equivalent number of sigma. The 511 keV data are from Weidenspointner et al. (2008).  }
\label{table:asym}
\vspace{-6mm}
\begin{center}
\begin{tabular}{|c|c|c|c|c|}
\hline
                                                         &  Negative                 & Positive                             &     Ratio   &   Significance   \\
                                                         & Longitude                 & Longitude                        &                 &   (sigma)  \\
\hline
511 keV disk ($cm^{-2}s^{-1}$)  & $4.3 \times 10^{-4}$ & $2.3 \times 10^{-4}$     &   	1.79    &  3.8   \\
  \hline
 3rd IBIS catalogue                        &   45                               &   26                                  &   1.73      &     2.39    \\
4th IBIS catalogue                        &   57                                &  39                                  &    1.46     &    1.74     \\
Liu et al. LMXB catalogue           &    94                               &   66                                 &  1.42      &    2.30      \\
Swift/BAT 58m survey        &   48                                 &  29                                 &  1.66      &   2.06       \\
\hline
\end{tabular}
\end{center}
\label{default}
\end{table}%

Recently Higdon et al. (2009) have developed an idea, similar to a suggestion of  Prantzos (2008), that  positrons produced in the disk are channeled by magnetic fields towards the bulge region, there to annihilate. They argue that  their detailed models of the propagation and annihilation of positrons produced by radioactive decay of $^{56}$Ni, $^{44}$Ti and $^{26}$Al can explain all of the observations, including the asymmetry.  A major uncertainty in such work is how to handle the positron transport in the presence of interactions with MHD waves that greatly affect the mean free path (Jean et al., 2009).

\section{Are  the high and low energy positrons  related?}
\label{link}

Although it is natural to consider the possibility that the high energy positrons are the source (or a major source) of the lower energy ones that lead to the observed annihilation radiation, there are several reasons to believe that this is  not the case.  Each line of argument is, however, subject to certain caveats.

\noindent {\bf Balancing the budget.}
The observed 511 keV flux of $\sim10^{-3}$ photons s$^{-1}$ cm$^{-2}$  implies 1--2$\times 10^{43}$  low energy positrons annihilate per second. Cosmic ray models predict a rate of production of high energy  positrons that is an order of magnitude lower. This argument assumes of course an equilibrium condition and that the cosmic ray models are reliable.

\noindent {\bf Escape.}
High energy positrons are likely to escape from the galaxy before slowing down to low energies. As positrons that initially have higher energies slow down to the GeV range, the characteristic time for further slowing becomes long ($\sim 10^9$ y) compared with typical escape times ($\sim 10^7$ y). Thus they are unlikely to reach the MeV regime, where energy loss time scales again become short.
Again the conclusion is dependent on modeling.

\noindent {\bf Absence of radiation while slowing.}
It has been argued that the positrons responsible for the 511 keV radiation cannot have originated with energies greater than a few MeV otherwise gamma-ray radiation produced at intermediate energies would exceed observed levels. Beacom et al. (2006) conclude that the initial energy cannot exceed  3 MeV, while on more conservative  assumptions Sizun et al. (2006)  place the limit at 7 MeV.  
Chernyshov et al. (2008)  conclude however that if the magnetic field in the galactic bulge is as high as a few milligauss, these limits do not apply.

Finally, as mentioned above, there are more than enough mechanisms that could produce positrons at low (MeV) energies without invoking another source.

\section{Limits on point sources}

The fact that the 511 keV emission appears `diffuse' with present instrumentation does not exclude the possibility  that it comprises 
many point sources or that point sources also contribute at some level. Searches for point sources made with the IBIS instrument on INTEGRAL, that has higher angular resolution than SPI (though poorer spectral resolution), place limits of a few times 10$^{-4}$ photons cm$^{-2}$ s$^{-1}$ ( De Cesare et al., 2009). Searches by Teegarden and Watanabe  (2006) and by Tsygankov and Churazov (2010)   failed to detect any variability in the 511 keV emission, that would be a clear indicator of point source contributions.  

\section{Alternatives to gamma-ray observations} 

An interesting possibility is that  `atomic'  transitions might allow  positronium  to be observed during the short `married' life of a positron and an electron before they annihilate. The energy level structure of such an atom is that of a hydrogen atom except that all the basic energies are a factor of 2 lower. The fine and hyperfine splitting are of course different as well -- for example the $1^1S_0 - 1^3S_0$  spin-flip transition, responsible for the H 21 cm line, appears at 1.47 mm.

Attempts have been made to observe some of the expected lines. Puxley and Skinner (1996) looked for  the Ps Paschen $\gamma$ line at 2.18 $\mu$m, while Anantharamaiah et al. (1989) searched for Ps radio recombination lines near 6 cm and 20 cm. Ellis and  Bland-Hawthorn  (2009) argue that on certain assumptions the recombination lines  should  be detectable with new technologies becoming available.

Observing of any of these lines 
would provide an independent way of investigating positrons in the galaxy but if the positronium distribution is entirely diffuse, then detection is likely to be problematic. 

\section{Conclusions and Prospects}

The prospects for improving our understanding of the origin of the low energy positrons are somewhat limited. INTEGRAL is continuing to operate and SPI observations are being made that are optimized to study the exact form of the emission from their annihilation, but as discussed above this may have limited relevance to where they are created. Furthermore the sensitivity and angular resolution of SPI are limited and no planned (or even seriously suggested) instrument can improve on its performance for diffuse line emission. For point or localized sources gamma-ray lenses could lead to an improvement in sensitivity but the incentive to develop such instrumentation is reduced by the absence of evidence that such sources exist.

\section{Acknowledgements}
The author wishes to thank
P. Jean, J, Kn\"odlseder, P. Martin, P. von Ballmoos and G. Weidenspointner
for valuable discussions.

\bigskip

\end{document}